\documentclass{aa}  

\usepackage{graphicx}
\usepackage{txfonts}
\usepackage[colorlinks, citecolor=blue]{hyperref}
\usepackage{natbib}

\begin{document}

   \title{Morphologies of galaxies within voids}

%   \subtitle{I. Overviewing the $\kappa$-mechanism}

   \author{M.~Argudo-Fernández\inst{1, 2} \and
          C.~Gómez Hernández\inst{1} \and
          S.~Verley\inst{1, 2} \and
          A.~Zurita\inst{1, 2}\and
          S.~Duarte Puertas\inst{1, 2, 3} \and
          G.~Blázquez Calero\inst{4} \and 
          J.~Domínguez-Gómez\inst{1} \and
          D.~Espada\inst{1, 2} \and
          E.~Florido\inst{1, 2} \and
          I.~Pérez\inst{1, 2} \and
          L.~Sánchez-Menguiano \inst{1, 2} 
          }

   \institute{Departamento de Física Teórica y del Cosmos, Edificio Mecenas, Campus Fuentenueva, Universidad de Granada, E-18071 Granada, Spain\\
              \email{margudo@ugr.es}
         \and
             Instituto Carlos I de Física Teórica y Computacional, Facultad de Ciencias, Universidad de Granada, E-18071 Granada, Spain
         \and 
         D\'epartement de Physique, de G\'enie Physique et d'Optique, Universit\'e Laval, and Centre de Recherche en Astrophysique du Qu\'ebec (CRAQ), Qu\'ebec, QC, G1V 0A6, Canada
        \and
         Instituto de Astrofísica de Andalucía - CSIC, Glorieta de la Astronomía s.n., E-18008 Granada, Spain
             }

   \date{Received May 21, 2024; accepted November 4, 2024}

% \abstract{}{}{}{}{} 
% 5 {} token are mandatory
 
  \abstract
   {Among the largest structures in which matter is distributed in the Universe, we can find large, underdense regions, almost devoid of galaxies, called cosmic voids. The study of these structures and the galaxies which inhabit them, the void galaxies, provide key information to understand galaxy evolution.}
   {In this work we investigate the effects of the environment on the evolution of void galaxies. In particular, we study their morphology and explore its dependence with their location within the void where the galaxies reside, as well as with properties of the void, such as void size or galaxy number-density.}
   {The sample of void galaxies that we use in this study is based on the catalogue of cosmic voids and void galaxies in the Sloan Digital Sky Survey Data Release 7 (SDSS-DR7). Since we are interested into study the morphology of void galaxies, we select galaxies in the redshift range 0.005\,$\leq$\,z\,$\leq$\,0.080, and use the public galaxy morphologies for SDSS with Deep Learning algorithms to divide the sample between early- and late-type void galaxies. We analyse the fraction of galaxies of each morphology type as a function of the void-centric distance, the size of the voids, and the  density of galaxies in each void.} 
   {There is a higher abundance of late-type galaxies with respect to early-type galaxies within voids, which remains nearly constant from the inner to the outer part of the voids. We do not find any dependence of the fraction of early- and late-type galaxies with respect to the size of the voids or the number-density of galaxies in the voids.} 
   {Galaxies in voids follow the morphology-density relation, in the sense that the majority of the galaxies in voids (the most under-dense large-scale environments) are late-type galaxies. However, we find no difference between voids with lower or higher volume number-density of galaxies: the fraction of early- and late-type galaxies do not depend on the density of the voids. The physical processes responsible for the evolution from late towards earlier types (such as external environmental quenching) are not sufficiently effective in voids or so slow (internal secular quenching) that their contributions do not appear in the morphology-density relation.}


   \maketitle
%
%-------------------------------------------------------------------

%__________________________________________________________________
%% 1. Introduction %%%%%%%%%%%%%%%%%%%%%%
\section{Introduction}

%%{\bf 1. Set the stage, address the title, frame the big  picture.Use the last sentence of the first paragraph to attack/defend/justify your research}
Cosmic voids are large underdense regions present in the hierarchical large scale structure of the universe. Surrounded by dense structure (filaments, walls, and clusters), voids are an essential component of the cosmic web \citep{1996Natur.380..603B,2011IJMPS...1...41V,2017ApJ...835..161M,2024MNRAS.527.4087J}, representing about the 80\% of the volume of the Universe. Cosmic voids have been recognised as interesting cosmological laboratories for investigating galaxy evolution, structure formation, and cosmology \citep{2007MNRAS.381...41H,2009MNRAS.395.1915T,2017MNRAS.470...85L,2013MNRAS.434.1435C,2024MNRAS.527.2663L}. In particular, there is an increased interest on galaxies residing in voids, the so-called void galaxies \citep{1999AJ....118.2561G,2002A&A...389..405P,2003MNRAS.340..160B,2004ApJ...617...50R}. Void galaxies are considered as the loneliest galaxies in the Universe, since they are in general further away from each other than galaxies that are located in clusters, filaments, or walls \citep{2020MNRAS.493..899H}. The pristine low-density environment of voids provides an ideal place to examine the influence of environment on the formation and evolution of galaxies. There are therefore many works that study the void galaxies and compare their properties with galaxies found in high-density regions in the Universe \citep[e.g.,][]{1996AJ....111.2150S,1997A&A...318..405K,1997A&A...325..881P,2005ApJ...620..618H,2006MNRAS.372.1710P,2015ApJ...798L..15K,2017MNRAS.464..666B,2022MNRAS.517..712R,2023MNRAS.521..916R}. The studies on void galaxies are raising some exciting questions and answers.

%%{\bf 2. Summarize what has been done up to present}
Generally, void galaxies contain less stellar mass, are bluer, have slightly lower stellar metallicities, and show later-type morphology than galaxies in denser environments \citep{2000AJ....119...32G,2004ApJ...617...50R,2005MNRAS.356.1155C,2010MNRAS.409..936P,2012MNRAS.426.3041H,2021ApJ...906...97F,2023MNRAS.524.5768P,2023A&A...680A.111D}. These differences are not easy to understand considering that, even although the fraction of star-forming galaxies in voids is higher than in filaments and walls \citep{2008MNRAS.390L...9C,2015ApJ...810..165L}, no significant differences are found in terms of luminosity, stellar populations, and specific star formation rates \citep{2015ApJ...815...40D,2016MNRAS.458..394B,2016ApJ...831..118M,2022A&A...658A.124D}. Contradictory results are found with respect to the gas content and gas-phase metallicities of void galaxies with respect to denser large-scale environments \citep{2011AstBu..66..255P,2015ApJ...798L..15K}. The discrepancies related to gas content and metallicites might be related to different star formation histories (SFHs), where void galaxies have, on average, slower SFHs than galaxies in filaments, walls, and clusters \citep{2023Natur.619..269D}. Moreover, these differences in SFH might be also related to the different properties of the voids, or to the location of the galaxies within voids. From numerical simulations, \cite{2024MNRAS.527.4087J} have shown that cosmic voids possess an intricate internal network of substructures, making them a complex environment for galaxy formation and evolution. Therefore, the internal matter distribution within the void (i.e. void environment) might have an impact on the properties and evolution of void galaxies. To understand the impact of the void environment on galaxy evolution it is important to have a well defined sample of void galaxies for a large and diverse sample of cosmic voids, covering the widest possible range of void properties. 

%%{\bf 3. State what is needed missing from the state of knowledge}
There are a few works that study properties of void galaxies, in particular, galaxy morphology, considering the properties of the void where galaxies reside. \citet{2017ApJ...846L...4R} explored the morphology of galaxies located in the proximity of 630 cosmic voids, selected from the catalogue of voids of \citet{2012ApJ...744...82V}, identified in the seventh data release of the Sloan Digital Sky Survey \citep[SDSS-DR7,][]{2009ApJS..182..543A}. Their sample of galaxies have a redshift range $0.01\,\leq\,z\,\leq\,0.12$, in a void size range between $10h^{-1}$\,Mpc and $18h^{-1}$\,Mpc. These galaxies are situated in the outskirts of voids (the void external region with $0.8<r/R_{void}<2.5$, where $r$ is the comoving distance from the center of a void and $R_{void}$ is the void radius). \citet{2017ApJ...846L...4R} found a significant correlation between the galaxy morphology and the distance from the center of the void, where the fraction of early-type galaxies increases, and the fraction of late-type galaxies decreases, when moving from the center to the edges of the voids. They also found that early-type galaxies tend to be found in the proximity of voids of a large size, therefore the predominant type or morphology of void galaxies is also influenced by the size of the void. However, it is important to note that \citet{2017ApJ...846L...4R} study focuses only on galaxies located in the surrounding of voids ($0.8<r/R_{void}<2.5$), not taking into account the galaxies located in the inner regions of the voids. Recently, \cite{2023MNRAS.521..916R} analysed whether the properties of void galaxies are sensitive to the type of voids that inhabit, as well as the local environment within the void, using information from the Sloan Digital Sky Survey Data Release 16 \citep[SDSS-DR16;][]{2020ApJS..249....3A}. They classified voids according to their integrated contrast density profile, separating in S and R type voids\footnote{R-type voids are those in expansion at all scales, while S-type voids are those expanding at their inner regions but collapsing at larger scales, see \citet{2013MNRAS.434.1435C} for further details.}, which is qualitatively consistent with the separation into small and large size in \citet{2017ApJ...846L...4R}, respectively. They found that, regardless of the void region (including the inner regions of the voids), galaxies in S-type voids are redder, passive, and more concentrated than galaxies in R-type voids. Their results support that the differences in the colour and morphology in the void outskirts might be related with their environment. The morphology classification used in \cite{2023MNRAS.521..916R} is based on concentration r$_{90}$/r$_{50}$ (where $_{90}$ and r$_{50}$ are the radii containing 90\% and 50\% of the total flux of the galaxy, respectively) which is only a proxy of the galaxy morphology. % and might lead to biased conclusions. 

%%{\bf 4. State the objectives of the paper}
In this work, we aim to better understand the effect of the void environment on the morphology of void galaxies, from the inner region to the edge of the voids. This work complements previous studies by characterising galaxies within voids, using a well defined sample of void galaxies in the local Universe. We use the \citet{2012MNRAS.421..926P} catalogue of cosmic voids and void galaxies to study how the galaxies are distributed inside voids according to their morphology, and how this distribution depends on the size of the voids. To account for the morphology of the galaxies, we use the morphology classification by \cite{2018MNRAS.476.3661D} for galaxies in the SDSS, who combined accurate existing visual classification catalogues with machine learning techniques, providing the largest and most accurate morphological catalogue up to date. %To compare with the results in \citet{2017ApJ...846L...4R} and \cite{2023MNRAS.521..916R}, we also use other morphology classifications from the literature to account for any possible bias. 

This work is organised as follows. In Section~\ref{sec:voidgal} we describe the sample of voids and void galaxies used in this work. In Section~\ref{sec:morpho} we describe the morphology classification method that we used for our sample. In Section~\ref{sec:res} we show the main results about the morphology of the void galaxies, and discuss these results in Section~\ref{sec:dis}, comparing with the ones obtained by \citet{2017ApJ...846L...4R} and \cite{2023MNRAS.521..916R}. Finally, in Section~\ref{sec:sum} we present the main conclusions of this work. Throughout the paper, we use a cosmology with $\Omega_{\Lambda 0} = 0.7$, $\Omega_{\rm m 0} = 0.3$, and $H_0 = 70$\,km\,s$^{-1}$\,Mpc$^{-1}$.

%__________________________________________________________________
%% 2. Data and samples %%%%%%%%%%%%%%%%%%%%%%
%\section{Data and samples}\label{sec:Data}

%In this section, we describe the cosmic void catalogues and the morphological information contained in the catalogues of galaxy morphologies.

\section{Data and the sample} \label{sec:data}

%__________________________________________________________________
%% 2.1. Void galaxy sample %%%%%%%%%%%%%%%%%%%%%%
\subsection{Void galaxy sample} \label{sec:voidgal}

The sample of void galaxies that we use in this study is based on the catalogues of cosmic voids and void galaxies compiled by \citet{2012MNRAS.421..926P} using photometric and spectroscopic data from the SDSS-DR7 \citep{2009ApJS..182..543A}.
\citet{2012MNRAS.421..926P} studied the distribution of cosmic voids and void galaxies using the three dimensional distribution of galaxies in the SDSS-DR7.%, brighter than an absolute magnitude limit of $M_{r}\,<\,-20.09$. 
\citet{2012MNRAS.421..926P} used the VoidFinder algorithm as described by \citet{2002ApJ...566..641H}, based on the original method by \citet{1997MNRAS.287..790E}. This is a galaxy-based void-finding algorithm that uses redshift data to objectively find voids using a nearest neighbour algorithm on a volume-limited galaxy catalogue. From this catalogue, galaxies are initially classified as wall or void galaxies\footnote{A void galaxy is a galaxy that lives in void regions, whereas wall galaxies live in filaments and clusters.}. If the distance to the third nearest neighbour is lower than 7\,Mpc, a galaxy is considered as a wall galaxy, and is considered as a potential void galaxy if the distance is larger \citep{2012MNRAS.421..926P}. Potential void galaxies are then removed from the initial sample, using the remaining galaxies (wall galaxies) to identify the voids. The VoidFinder algorithm \citep{1997MNRAS.287..790E, 2002ApJ...566..641H} grows empty cells as spheres in the distribution of wall galaxies, merging the spheres when there is more than 50\% overlap. The centre of the void is also defined by the algorithm, however, the centre of the maximal sphere is not confined to the initial cell, but when the sphere is bound by four wall galaxies, with a cut-off of 10\,$h^{-1}$\,Mpc for the minimum radius of a void region \citep{2012MNRAS.421..926P}. Finally, potential void galaxies are put back to identify the ones falling within voids as void galaxies. In this context, the SDSS main galaxy sample (with redshift completeness for galaxies with Petrosian $r$-band magnitude $r$\,<\,17.77) is ideal to identify a statistically significant sample of cosmic voids. 

\citet{2012MNRAS.421..926P} used the Korea Institute for Advanced Study Value-Added Galaxy Catalogue \citep[KIAS-VAGC;][]{2010JKAS...43..191C}, which main source is the New York University Value-Added Galaxy Catalogue large-scale structure sample \citep[NYU-VAGC;][]{2005AJ....129.2562B}, which is also based on the SDSS-DR7. This resulted in a primary sample of 120606 galaxies, in a magnitude range with high completeness in the range 10.0\,$\leq$\,$r$\,$\leq$\,17.6, in a volume-limited sample with redshift $z$\,<\,0.107 and absolute magnitude in the r-band $M_r$\,<\,$-$20.09. Using this sample, they identified 1054 statistically significant voids in the northern galactic hemisphere greater than 10\,$h^{-1}$\,Mpc in radius, covering 62\% of the volume of the Universe up to that redshift range. These voids are populated by 8046 galaxies brighter than an absolute magnitude limit of $M_{r}\,<\,-20.09$ (the volume-limited galaxy sample) and 79947 void galaxies with $r$\,$\leq$\,17.6 and $z$\,<\,0.088 (the magnitude-limited galaxy sample).

\begin{figure*}
\begin{center}
% [trim={left bottom right top},clip]
\includegraphics[width=\textwidth]{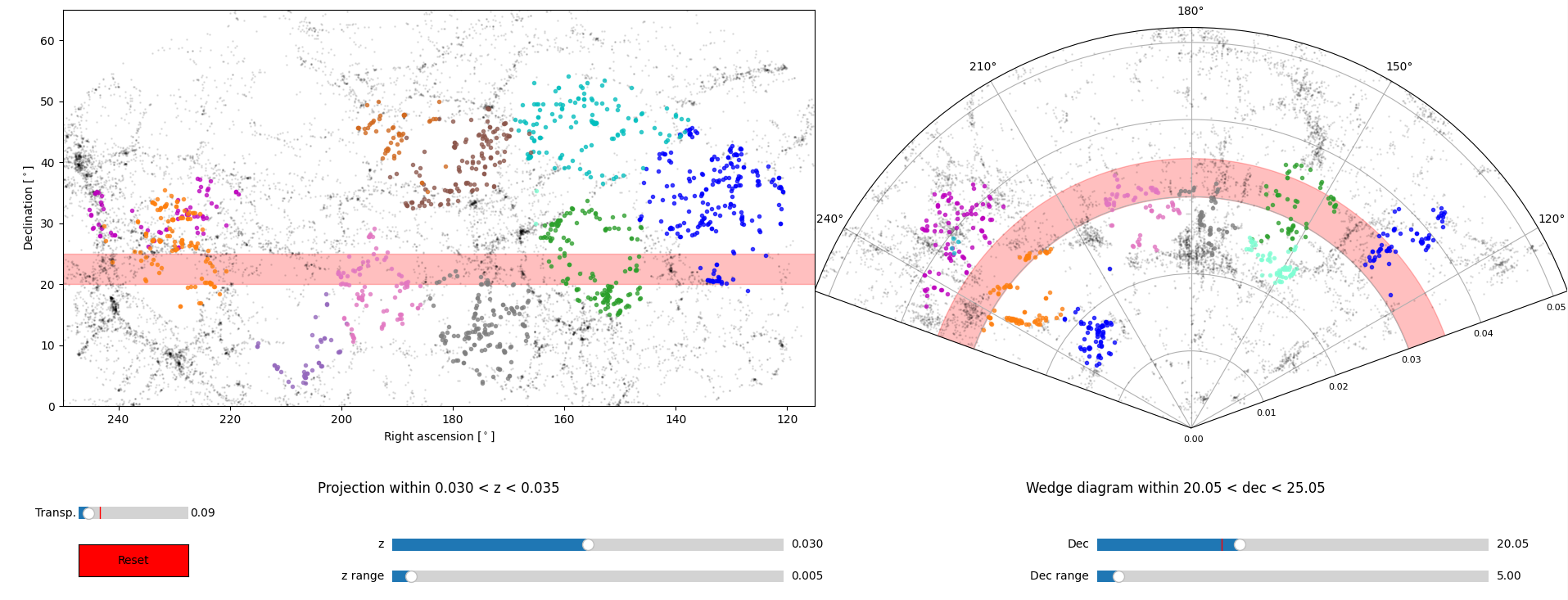}
\caption{Spatial distribution of voids in the Voids174 sample using an updated version of the LSSGalPy interactive 3D visualisation tools \citep[][Alcázar-Laynez et al. in prep.]{2017PASP..129e8005A} publicly available at \url{https://gitlab.com/astrogal}. Only the voids within the redshift range $0.030 \leq $z$ \leq 0.035$ (left-hand panel) and within $\rm 20.05\,deg \leq dec \leq 25.05\,deg$ (right-hand panel) are shown. Individual voids are represented in different colour. Galaxies in the large-scale structure in the same redshift and declination range are represented by black points. To guide the eye, the red stripe in the left panel shows the selected declination range in the wedge diagram, and the selected redshift range in the ra-dec projection corresponding to the right panel.}
\label{fig:lavin}
\end{center}
\end{figure*}

For the purpose of this study, we use the magnitude-limited sample, to have a sample size sufficiently robust, and applied a volume-limited sample selection to check and correct any observing bias in the results. Hereafter, we will refer to the magnitude-limited galaxy sample as the Pan+12 sample. Since we are interested on studying the morphology of void galaxies, we select galaxies in the redshift range 0.005\,$\leq$\,$z$\,$\leq$\,0.080 to reduce uncertainties in the visual morphological classification, especially for fainter and/or smaller galaxies, due to poor spatial resolution in SDSS images. This leaves us with 26864 void galaxies in 174 voids (11624 void galaxies in the volume-limited sample with $M_r$\,<\,$-$20.09 up to $z$\,=\,0.08). Hereafter, we will refer to this as the Voids174 sample. Figure~\ref{fig:lavin} shows the spatial distribution of the voids in the Voids174 sample, where the void region is well delimited by denser large-scale structures, as filaments and walls, with clusters found in the intersection of these. 

\begin{figure*}
\begin{center}
% [trim={left bottom right top},clip]
\includegraphics[width=0.48\textwidth, trim={0 0 1cm 0},clip]{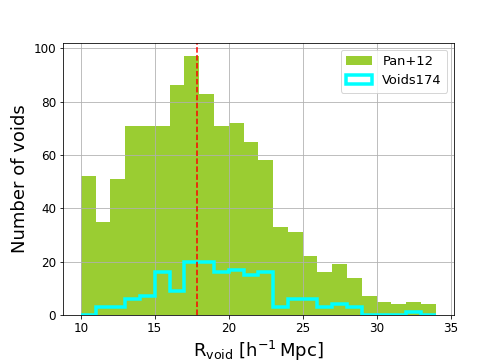}
\includegraphics[width=0.48\textwidth, trim={0 0 1cm 0},clip]{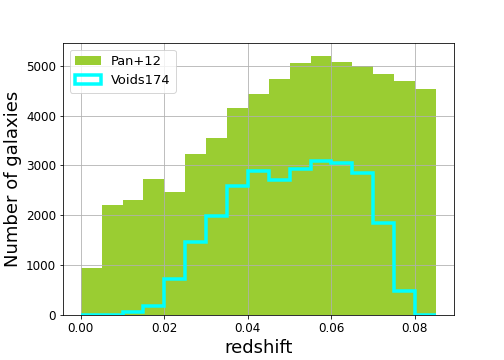}
\caption{Left panel: Relative number of voids as a function of the void size ($\rm R_{void}$), measured by the effective radius, in units of $\rm h^{-1}\,Mpc$. The green histogram represents the distribution for voids in the Pan+12 sample (1055 voids) and the cyan histogram represents the distribution for voids in the Voids174 sample (174 voids). The dashed red vertical line indicates the median value of the effective radius for the Pan+12 sample ($17.83h^{-1}$\,Mpc). Right panel: Distribution of voids galaxies as a function of redshift. The green histogram represents the distribution for void galaxies in the Pan+12 sample (79947 galaxies) and the cyan histogram represents the distribution for void galaxies in the Voids174 sample (26864 galaxies).}
\label{fig:reff_z}
\end{center}
\end{figure*}

The size of the voids ($R_{void}$) is measured using the effective radius ($r_{eff}$). The effective radius represents the radius of voids when they are spherical or, in the case where the void is not spherical, the radius of the sphere that has the same volume as the non-spherical void. The voids have a median effective radius of $17.83\,h^{-1}$\,Mpc, with values ranging from $10\,h^{-1}$\,Mpc to $30\,h^{-1}$\,Mpc. We use the median value of the effective radius in the Pan+12 sample ($17.83\,h^{-1}$\,Mpc) to separate our sample into small (61 voids, containing 6285 void galaxies) and large (113 voids, containing 20579 void galaxies) voids. This value is also in agreement with \citet{2002ApJ...566..641H}, who claim that voids have an average effective diameter of 29.5\,$h^{-1}$\,Mpc ($\sim$14.8\,$h^{-1}$\,Mpc effective radius), and \cite{2023MNRAS.521..916R}, who identified large voids when their effective radius is larger than 18\,$h^{-1}$\,Mpc. The left and right panels of Fig.~\ref{fig:reff_z} show the distributions of void effective radius and redshift for the void and void galaxies in the Voids174 sample, respectively. The same distributions for the Pan+12 sample are shown in green for comparison. The amount of voids increases at higher redshift because the comoving volume spanned by the survey is larger. 

\begin{figure}
\begin{center}
% [trim={left bottom right top},clip]
\includegraphics[width=\columnwidth, trim={0 0 1cm 1cm},clip]{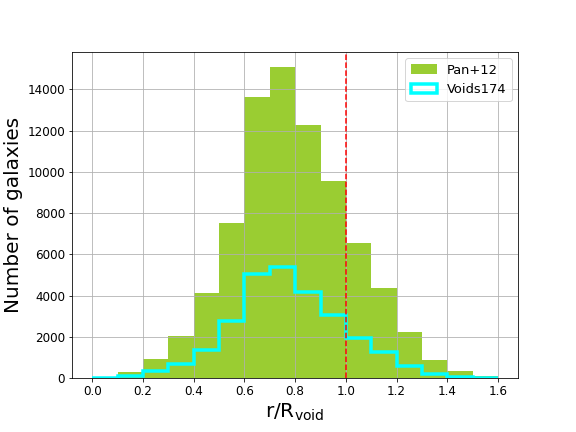}
\caption{Comparison between the normalised distance with respect to the effective radius of each void ($r/R_{void}$) for void galaxies in \citet{2012MNRAS.421..926P} (79947 galaxies, green histogram) and the subsample of void galaxies considered in this study (26864 galaxies, cyan histogram). The dashed red vertical line represents the edge of the voids if they were spherical.}
\label{fig:normreff}
\end{center}
\end{figure}

In order to be able to compare the properties of the galaxies with respect to their position within the voids, we use the normalised distance from the void centre, by dividing the void-centric distance by the size of the void $r/R_{void}$. Voids are in general spherical but some of them may present a different shape (e.g., elliptical, ovoid, elongated), therefore, the value $r/R_{void}$ can be greater than one for the galaxies located outside the sphere with radius $r_{eff}$, as shown in Fig.~\ref{fig:normreff}. Note that the parameter $r/R_{void}$ can potentially mix void galaxies that may belong to different internal void substructures (tendrils, groups, slight local overdensities) and/or may be closer or farther from the surrounding wall, filament or cluster. Nevertheless, the void galaxy number-density is constant and does not depend on $r/R_{void}$ within the void environment, therefore the results in the present work hold for both the most spherical and also the potentially less spherical voids, independently of the void shape.

The catalogue of voids in the Pan+12 sample does not contain information about the morphology of the void galaxies, so we complete this information from the literature as described in Sect.~\ref{sec:morpho}.

% Figura 1__________________________________________________________________
% __________________________________________________________________

% Figura 2__________________________________________________________________
% __________________________________________________________________

% Figura 3__________________________________________________________________
% __________________________________________________________________

%__________________________________________________________________
%% 2.2. Galaxy morphology %%%%%%%%%%%%%%%%%%%%%%
\subsection{Galaxy morphology} \label{sec:morpho}

Since the Pan+12 sample is based on SDSS data, we use the morphology classification by \citet{2018MNRAS.476.3661D} (hereafter DS18), which provides the largest and most accurate morphological catalogue up to date for SDSS galaxies. In particular, they provided deep learning-based morphological classification for $\sim$670\,000 galaxies in the SDSS-DR7. The models used in \citet{2018MNRAS.476.3661D} were trained and tested using accurate existing visual classification catalogues from the \textit{Galaxy Zoo} project,\footnote{\texttt{\url{www.galaxyzoo.org}}} in particular from the Galaxy Zoo 2 \citep[GZ2,][]{2013MNRAS.435.2835W} catalogue and the morphological classification by \citet{2010ApJS..186..427N}.

The morphology of the galaxies in the DS18 catalogue is given through the numerical parameter 'T-Type', which ranges from $-$3 to 10, where T-Type~$\leq$~0 corresponds to early-type galaxies (i.e., E and S0), and positive values to late-type galaxies (i.e., Sa, Sb, Sc, Sd, and Sm), with T-Type~=~10 for irregular galaxies\footnote{Note that is not trivial to relate the T-Type parameter, as defined in DS18, with the historical Hubble type morphological definition for spiral galaxies (Sa, Sb, Sc, Sd, and Sm), considering than less than 1\% of galaxies in DS18 have T-Type\,>\,6.}. The T-Type is a continuous morphology parameter related to the Hubble sequence (Hubble T), and the GZ2 classification, following the \cite{1963ApJS....8...31D} morphology classification scheme. These correlations facilitate comparisons with different morphological classifications, and henceforth, with previous studies.

\begin{figure}
\begin{center}
% [trim={left bottom right top},clip]
\includegraphics[width=\columnwidth, trim={0 0 1cm 1cm},clip]{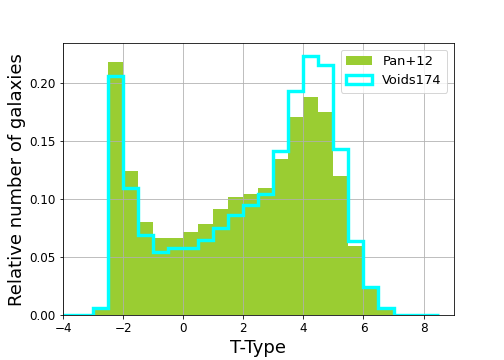}
\caption{Comparison between the T-Type morphological parameter in the DS18 catalogue for void galaxies in \citet{2012MNRAS.421..926P} (72118 galaxies, green histogram) and the subsample of void galaxies in this study (26864 galaxies, cyan histogram). Negative values (T-type $\leq$ 0) correspond to early-type galaxies while positive values (T-type $>$ 0) correspond to late-type galaxies.}
\label{fig:DS18all}
\end{center}
\end{figure}

The DS18 catalogue contains morphological classification for 72118 void galaxies in the Pan+12 catalogue (90\% of the sample\footnote{Note that the DS18 catalogue does not contain morphological classification for all the galaxies in the SDSS but for a large fraction of the SDSS spectroscopic main galaxy sample.}), while the GZ2 catalogue only contains information for 41\% of the Pan+12 catalogue. When we restrict to our sample of 174 voids, there is morphology classification for 25174 void galaxies out of 26864 (94\% of the sample). As shown in Fig.~\ref{fig:DS18all}, the distribution of the T-Type parameter between the Pan+12 and the Voids174 sample is similar, following a bimodal distribution for early- and late-type galaxies, however the Voids174 sample contains slightly more late-type galaxies, which is expected since the number of observable late-type galaxies decreases with redshift in magnitude limited samples \citep[the redshift completeness of the SDSS main spectroscopic sample is $m_{r,Petrosian}$\,<\,17.77\,mag;][]{2002AJ....124.1810S}.

%\begin{table}[h]
%\centering
%\begin{tabular}{ c  c  c }
%  \hline\hline  
%  (1) & (2) & (3) \\
% T-Type (DS18) & Hubble T & de Vaucouleurs class \\ 
% \hline
%  $[-3.0, -0.5)$ & $<$ 0 & Elliptical \\
%  $[-0.5, 0.5)$ & 0 & Lenticular \\
%  $[0.5, 2.5)$ & 1 & Sa \\
%  $[2.5, 4.5)$ & 3 & Sb  \\
%  $[4.5, 6.5)$ & 5 & Sc  \\
%  $[6.5, 9)$ & 7 & Sd  \\  
%  $-$ & 10 & Irregular  \\ 
%  \hline
%\end{tabular}
%\caption[]{\textcolor{red}{Duda, quizás se quite porque confunde.}The columns correspond to: (1) morphological classification according to the T-Type parameter in the DS18 catalogue; (2) Values of the numerical Hubble stage T; (3) corresponding \citet{1963ApJS....8...31D} morphological type.}
%\label{tab:DS18}
%\end{table}

% Figura 4__________________________________________________________________
% __________________________________________________________________

%__________________________________________________________________
%% 3. Results %%%%%%%%%%%%%%%%%%%%%%

\section{Results}\label{sec:res}

In this work we investigate the effect of the Universe large scale environment on the evolution of galaxies, by analysing the morphology of galaxies inhabiting cosmic voids. In particular, we study the morphology of galaxies as a function of the void-centric distance and the void size. In this section we present the results of the work. In Sect.~\ref{sec:res-galmorpho} we analyse the morphological classification of void galaxies, and compare with other classifications available in the literature. In Sect.~\ref{sec:res-galmorphodistance} and Sect.~\ref{sec:res-galmorphosize} we study the distribution of different galaxy morphological types as a function of the distance from the center of the void and the size of the void, respectively. 

%__________________________________________________________________
%% 3.1. Morphology of void galaxies %%%%%%%%%%%%%%%%%%%%%%
\subsection{Morphology of void galaxies}\label{sec:res-galmorpho}

As previously introduced, there is morphology information from the DS18 catalogue for 25174 void galaxies of 26864 ($\sim$\,94\% of the galaxies in the Voids174 sample). Of them, all the galaxies can be classified into early- or late-type galaxies using the T-Type parameter (as shown in Fig.~\ref{fig:DS18all}). We found that 6335 void galaxies (25.2\%) are classified as early-type (T-Type\,$\leq$\,0), while 18839 void galaxies (74.8\%) are classified as late-type (T-Type\,>\,0). The fraction of early- and late-type galaxies ($\rm f_{Early-type}$ and $\rm f_{Late-type}$, respectively) are calculated considering the the number of early- and late-type galaxies (as explained in Sect.~\ref{sec:morpho}) with respect to the total number of void galaxies with morphological classification ($N_T$). Uncertainties are given considering a binomial distribution as $e_f~=~\sqrt{\frac{f(1-f)}{N_T}}$.

%\begin{equation}
%\rm f_{Early-type} = N_{Early-type}/N_T\quad,
%\label{eq:fE}
%\end{equation}
%and 
%\begin{equation}
%\rm f_{Late-type} = N_{Late-type}/N_T\quad,
%\label{eq:fS}
%\end{equation} 
%where $\rm N_{Early-type}$ and $\rm N_{Late-type}$ are the number of early- and late-type galaxies (as explained in Sect.~\ref{sec:morpho}), respectively, and $\rm N_T$ is the total number of void galaxies with morphological classification.

Table~\ref{tab:morph-cat} shows comparison numbers using other galaxy morphology classifications, such as the morphological information from the GZ2 catalogue, which contains detailed morphology information for SDSS galaxies, as well as its first version, Galaxy Zoo 1 \citep[GZ1][]{2008MNRAS.389.1179L}, which is the catalogue used by \citet{2017ApJ...846L...4R}. In addition, we computed the concentration index (CI\,=\,r$_{90}$/r$_{50}$ in the $r$-band, using photometric information for void galaxies from public SDSS data) for the void galaxies in the Voids174 sample to compare our results with \cite{2023MNRAS.521..916R}. The CI parameter correlates with galaxy morphology, where galaxies that are classified as early-type (E, S0, and Sa) by visual classification methods usually have values of CI\,>\,2.6, and lower values for late-type galaxies \citep{2001AJ....122.1861S}. This parameter can be estimated for a large number of galaxies, except for some galaxies with flags in their r$_{90}$ and/or r$_{50}$ values (130 galaxies), however it is only a proxy for galaxy morphology and may lead to biases in the interpretation of the results. In fact, the fraction of galaxies classified as early-type using the CI parameter is higher than using the DS18 catalogue ($\sim$\,7\% higher), which might be related to the fact that early-type galaxies using the CI value also include disk galaxies (Sa), while the DS18 classification only includes E and S0 galaxies in this category. To evaluate the consistency with different morphology classification when interpreting our results, we have estimated the percentage of galaxies with the same morphology as in the DS18 catalogue (presented in the last column of Table~\ref{tab:morph-cat}). Even if the number of galaxies with GZ1 classification is the lowest, the classification between early- and late-type galaxies is the most consistent with the DS18 classification. As expected, the percentage of galaxies classified as early-type using the CI parameter and also classified as early-type in the DS18 catalogue is the lowest (about 51\%).

% Table 1__________________________________________________________________
\begin{table*}
\begin{center}
\begin{tabular}{c c c c c c}
\hline\hline
 (1) & (2) & (3) & (4) & (5)\\
 Source & Classified (\%) & Early-type (\%) & Late-type (\%) & In DS18 \\ 
 \hline
DS18 & 25174 (93.7) & 6335 (25.2\,$\pm$\,0.3) & 18839 (74.8\,$\pm$\,0.3) & -- \\ 
GZ1 & 9612 (35.8) & 815 (8.5\,$\pm$\,0.3) & 8797 (91.5\,$\pm$\,0.3) & 93.9\,$\pm$\,0.8 | 94.3\,$\pm$\,0.2 \\
GZ2 & 13487 (50.2) & 4014 (29.7\,$\pm$\,0.4) & 9473 (70.2\,$\pm$\,0.4) & 66.8\,$\pm$\,0.7 | 84.9\,$\pm$\,0.4 \\
CI & 26734 (99.5) & 8726 (32.6\,$\pm$\,0.3) & 18008 (67.4\,$\pm$\,0.3) & 51.4\,$\pm$\,0.5 | 84.9\,$\pm$\,0.3 \\ 
\hline
\end{tabular}
\caption{Number of void galaxies in the Voids174 sample with morphological classification using different catalogues: DS18 catalogue, Galaxy Zoo 1 (GZ1), Galaxy Zoo 2 (GZ2), and using the concentration index (CI) computed from photometric SDSS data. The columns correspond to: (1) catalogue or method used to classify galaxy morphology; (2) Number of void galaxies with morphological classification and corresponding percentage with respect to the total number of void galaxies in the Voids174 sample (26864 galaxies); (3) Number of galaxies classified as early-type in the corresponding catalogue and percentage of galaxies with respect to column (2), i.e., fraction of early-type galaxies ($\rm f_{Early-type}$); (4) same as (3) for late-type galaxies, with the corresponding fraction of late-type galaxies ($\rm f_{Late-type}$). Uncertainties are given considering a binomial distribution as $e_f~=~\sqrt{\frac{f(1-f)}{N_T}}$, where $f$ is the relative fraction for the total number of $N_T$ galaxies classified in each sample shown in column (2) ; and (5) Fraction of galaxies with early/late type morphology in common with the DS18 morphology classification, respectively. In this case uncertainties are estimated considering $N_T$ as the total number of galaxies with early/late type morphology on each catalogue shown in columns (3) and (4), respectively.}
\label{tab:morph-cat}
\end{center}
\end{table*} 
% __________________________________________________________________
%\sqrt(f (1-f)/N_T)

%__________________________________________________________________
%% 3.2. Morphology of void galaxies as a function of the void-centric distance %%%%%%%%%%%%%%%%%%%%%%
\subsection{Morphology of void galaxies as a function of the void-centric distance}\label{sec:res-galmorphodistance}

\begin{figure*}
\begin{center}
% [trim={left bottom right top},clip]
\includegraphics[width=\textwidth]{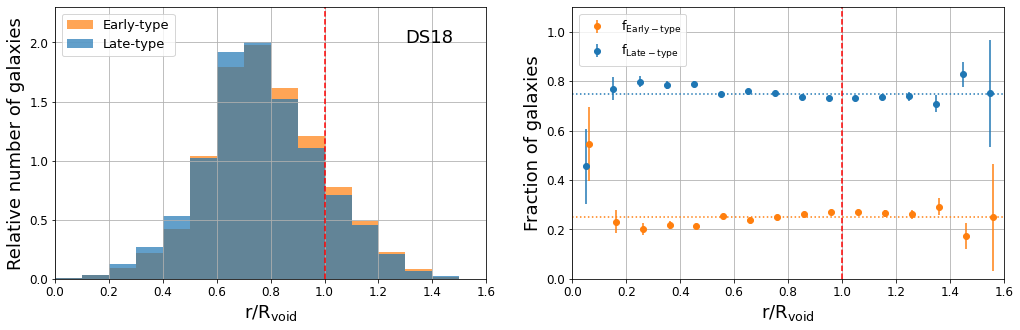}
\caption{Morphology of void galaxies with respect to the normalised effective radius of the voids. The dashed red vertical line represents the edge of the voids if they were spherical. Left: Relative number of early- (6225 galaxies, orange histogram) and late-type (18839 galaxies, blue histogram) void galaxies. Right: Fraction of early- and late-type galaxies, with their correspondending errors. Early-type galaxies are shown in orange and late-type galaxies in blue. Orange and blue dotted horizontal lines indicate the mean fraction of early- and late-type galaxies, 25.2\% and 74.8\%, respectively, as presented in Table~\ref{tab:morph-cat}. We use bins of 0.10\,$r/R_{void}$. The fraction of early-type galaxies has been shifted by 0.01\,$r/R_{void}$ so that the error bars do not overlap.}
\label{fig:morpdist}
\end{center}
\end{figure*}

In this subsection we study how galaxies are distributed according to their morphology in low-density environments, that is, in cosmic voids. In Fig.~\ref{fig:morpdist}, we present the distribution of galaxies in the inner region of the voids as a function of their morphology according to the DS18 classification. We study this behaviour for the galaxies in the Voids174 sample, separating the sample in early- and late-type galaxies. In the left-hand panel of the figure we show the distribution of early- and late-type galaxies as a function of the distance from the void centre, normalised to the void effective radius. In the right-hand panel we show the fraction of early-type ($\rm f_{Early-type}$) and late-type ($\rm f_{Late-type}$) galaxies, as defined in Sect.~\ref{sec:res-galmorpho}, following the same colour scheme. 
%The area shaded in grey represents the range of void-centric distances studied by \citet{2017ApJ...846L...4R}, which only takes into account the galaxies that are situated at the edges and surroundings of the voids. 
We see that, in general, the $\rm f_{Late-type}$ is higher than the $\rm f_{Early-type}$, and both fractions are nearly constant, with no significant differences, independently of the location of the galaxies within the voids. We discuss these results in Sect.~\ref{sec:dis-galmorphodistance}.

% Figura 5__________________________________________________________________
% __________________________________________________________________

%__________________________________________________________________
%% 3.3. Morphology of void galaxies as a function of the void size %%%%%%%%%%%%%%%%%%%%%%
\subsection{Morphology of void galaxies as a function of the void size}\label{sec:res-galmorphosize}

The last result of our work comes from the study of how galaxies with different morphology are distributed in large and small voids. We have divided the Voids174 sample into two subsamples of small and large voids. We have used as discriminant size the median value of the void size for the Pan+12 sample ($17.83\,h^{-1}$\,Mpc). Voids with a smaller void size are considered as small voids and those with a larger void size, as large voids. The number of voids, and the total number of galaxies contained in small and large voids are summarised in Table~\ref{tab:void-size}, including the number of early- and late-type galaxies, according to their classification in the DS18 catalogue. 

\begin{figure}
\centering
% [trim={left bottom right top},clip]
%\includegraphics[width=1\textwidth, trim={1cm 0 1cm 1cm},clip]{Figures/frac_early_sizevoids_174_240430.png}\\
%\includegraphics[width=1\textwidth, trim={1cm 0 1cm 1cm},clip]{Figures/frac_late_sizevoids_174_240430.png}
\includegraphics[width=1\columnwidth, trim={0 0 1cm 1cm},clip]{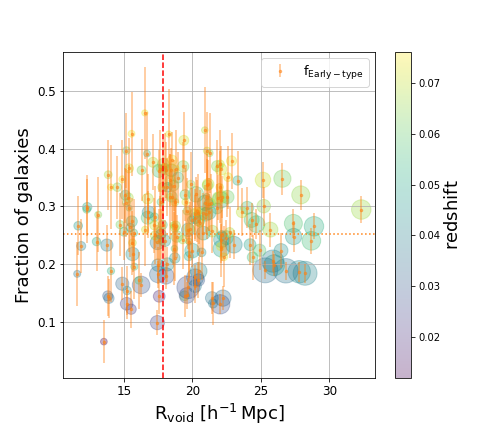}\\
\includegraphics[width=1\columnwidth, trim={0 0 1cm 1cm},clip]{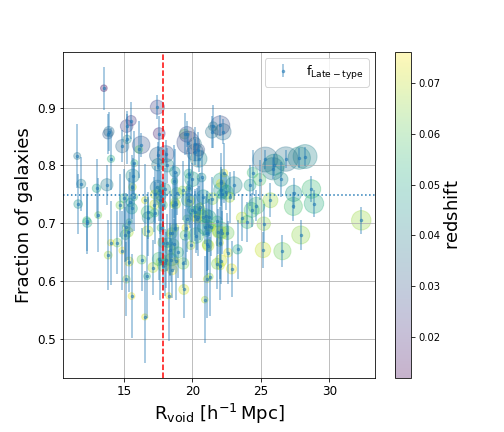}
\caption{Fraction of early-type and late-type galaxies in each void as a function of its size (measured by its effective radius). Each data point corresponds to a single void in the Voids174 sample, in the upper panel for the fraction of early-type galaxies, and in the lower panel for the fraction of late-type galaxies, thus for the same void $\rm f_{Early-type}$ + $\rm f_{Late-type}$ = 1. The size of the circle around each data point is proportional to the number of galaxies in each void, with a minimum of 37 galaxies and a maximum of 622 galaxies. The colour of each circle corresponds to the redshift of the void according to the colourbar at the right side of each panel. The red dashed vertical line represents the median value of void sizes $17.8 h^{-1}$ Mpc; so that small voids are located in the left-hand side of the plot and large voids towards the right-hand side. Orange and blue dotted horizontal lines indicate the mean fraction of early- and late-type galaxies, 25.2\% and 74.8\% in the Voids174 sample, respectively, as presented in Table~\ref{tab:morph-cat}. }
\label{fig:void-size}
\end{figure}

Additionally, Fig.~\ref{fig:void-size} shows the fraction of early-type galaxies ($\rm f_{Early-type}$, upper panel) and the fraction of late-type galaxies ($\rm f_{Late-type}$, lower panel) in each void, as a function of its void size. Void size ranges from $9.85\,h^{-1}$\,Mpc to $33.92\,h^{-1}$\,Mpc. The dashed-red vertical line marks the small voids from the large ones. Thus, we can use this information to explore the abundance of early- and late-galaxies in both type of voids and compare them. In general we do not observe any trend with the void size but there is a large scatter, which has a dependence with the redshift of the void. At a given void size, the $\rm f_{Early-type}$ increases with redshift, which is expected considering the magnitude-dependent Malmquist bias \citep{2005MNRAS.362..321B} implicit in the SDSS galaxy sample limited to $m_{r,Petrosian}$\,<\,17.77\,mag, which leads to the preferential detection of intrinsically bright objects, preferentially early-type galaxies. We have checked that the fractions of early- and late-type galaxies are still nearly constant with the void size at different redshift bins and also considering an absolute magnitude volume-limited sample. Note that the void effective radius in \citet{2012MNRAS.421..926P} is defined as the radius of the sphere of equivalent volume of the void. Therefore, the results of this study do not depend of the shape of the void. We discuss our results in Sect.\ref{sec:dis-galmorphosize}.

% Tabla 2__________________________________________________________________
\begin{table*}
\centering
\begin{tabular}{cccc}
\hline \hline
(1) & (2) & (3) & (4) \\
 sample & subsample & small voids & large voids \\ 
\hline
Voids & -- &61 & 113 \\ 
\hline
Void galaxies  & all & 6285 & 20579 \\
 & DS18 & 5846 & 19328 \\
 & Early-type & 1431 (24.5\,$\pm$\,0.6) & 4904 (25.4\,$\pm$\,0.3)\\
 & Late-type & 4415 (75.5\,$\pm$\,0.6) & 14424 (74.6\,$\pm$\,0.3)\\
\hline
\end{tabular}
\caption{Number of small and large voids in the Voids174 sample (upper row) and number of void galaxies in each void size (lower rows). For void galaxies with morphological classification in the DS18 catalogue, we also present the number (and fraction in percentage) of galaxies classified as early- and late-type for each void size. Fractions and uncertainties are estimated as indicated in Table~\ref{tab:morph-cat}.}
\label{tab:void-size}
\end{table*}
% __________________________________________________________________

% Figura 6__________________________________________________________________
% __________________________________________________________________

%__________________________________________________________________
%% 5. Discussion %%%%%%%%%%%%%%%%%%%%%%

\section{Discussion}\label{sec:dis}

\subsection{Morphology of void galaxies}\label{sec:dis-galmorpho}

As expected the distribution of the morphology T-Type parameter shows a bimodal distribution (as shown in Fig.~\ref{fig:DS18all}). Most of the galaxies in voids are late-type (Sa, Sb, Sc, Sd), with T-Type positive values ($\sim$\,75\%), and a smaller fraction are early-type (E and S0) galaxies, with T-Type~$\leq$~0 ($\sim$\,25\%). We obtain consistent results when using other morphology classifications (GZ2 and concentration index, see Table~\ref{tab:morph-cat}). The four morphological classifications show that there are more spiral galaxies than elliptical galaxies within voids. Note that the fraction of early-type galaxies using the GZ1 catalogue is much lower than the fraction of late-type galaxies, which might be due to the low completeness of the sample (less than 36\% of the sample has morphological classification). Therefore, in agreement with previous studies, late-type is the predominant morphology in voids \citep{2004ApJ...617...50R,2017ApJ...846L...4R,2021ApJ...906...97F,2024arXiv240410823C}. Note that due to Malmquist bias, the fraction of late-type galaxies decreases with the redshift, because only the brightest galaxies are observed. When considering the volume-limited sample of void galaxies, we find that the fraction of late-type galaxies ($\sim$\,64\%) is also larger than the fraction of early-type galaxies ($\sim$\,36\%), and therefore the results are not affected by this observing bias. 

The morphology of galaxies in the Universe depends on the density of the environment: the low-density regions, known as voids, present a higher abundance of late-type galaxies than the high-density areas, such as clusters and filaments. In fact, $\sim$\,62\% of galaxies in the DS18 sample (within the same redshift range as galaxies in the Voids174 sample, and removing the void galaxies) are late-type galaxies, with respect to $\sim$\,75\% in voids, and $\sim$\,38\% are early-type galaxies, with respect to $\sim$\,25\% in voids. Therefore, there is a higher fraction of late-type galaxies in voids than in denser environments (filaments, walls, and clusters). This result is consistent with the morphology-density relation \citep{1974ApJ...194....1O,1980ApJ...236..351D,1984ApJ...281...95P,2003MNRAS.346..601G,2010ApJ...714.1779V,2011ASSP...27....1D,2015MNRAS.451.3427H,2023MNRAS.518.5260P}.

\subsection{Morphology of void galaxies as a function of the void-centric distance}\label{sec:dis-galmorphodistance}

Although the number of late-type galaxies is larger than the number of early-type galaxies, their distribution as a function of the normalised void-centric distance is similar. As shown in the left panel of Fig.~\ref{fig:morpdist}, the total number of galaxies increases from the inner region of the voids up to a distance of $\rm r/R_{void}$\,$\simeq$\,0.75, from where the number of void galaxies decreases down to zero in the outer regions of the voids. Despite of this general trend, the fraction of early- and late-type galaxies is nearly constant up to a normalised distance of $\rm r/R_{void}\,\lesssim\,1.3$, with median $\rm f_{Late-type}$\,$\simeq$\,75\% and median $\rm f_{Early-type}$\,$\simeq$\,25\% (as shown in the right panel of Fig.~\ref{fig:morpdist}). There is a large dispersion with high error bars in the very inner and outer regions of the voids. This is due to the geometric effect that occurs in the centre of voids where we find few galaxies because of the small volume of the void that we consider (normalised distance of $\sim0.1$) and for the geometric effect in the edge of the voids due to the non-spherical shape (e.g. peanut-shape) of some voids (normalised distance >\,1.2). %These high error values make the data points less reliable at those void-centric distances. 

\begin{figure*}
\begin{center}
% [trim={left bottom right top},clip]
\includegraphics[width=\textwidth]{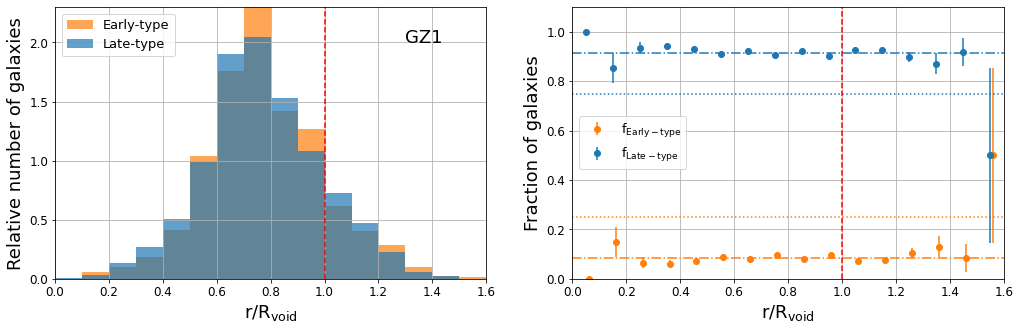}\\
\includegraphics[width=\textwidth]{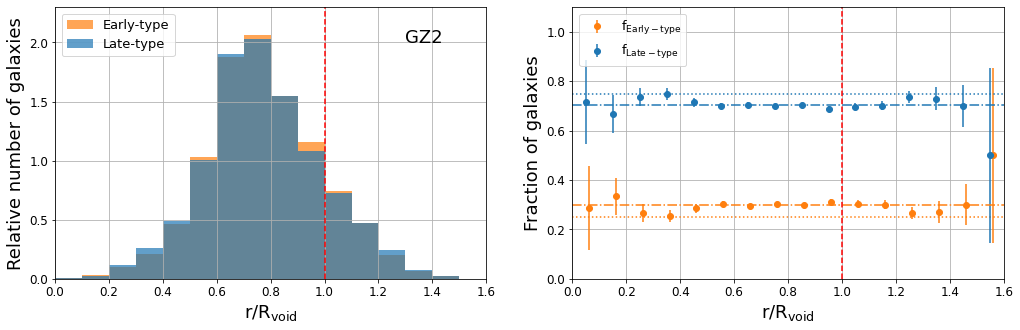}\\
\includegraphics[width=\textwidth]{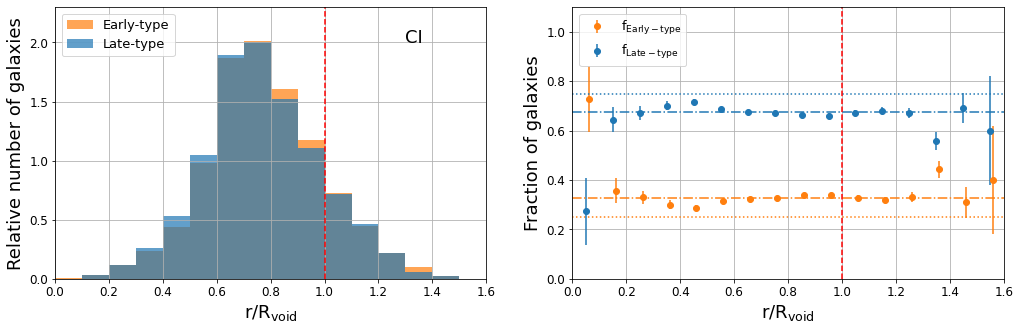}
\caption{Similarly as Fig.~\ref{fig:morpdist}, morphology of void galaxies with respect to the normalised effective radius of the voids, considering the GZ1 morphology catalogue (upper panel), the GZ2 morphology catalogue (middle panel), and the CI parameter (lower panel). The dashed red vertical line represents the edge of the voids if they were spherical. Left panels: Relative number of early- (orange histograms) and late-type (blue histograms) void galaxies. Right: Fraction of early- and late-type galaxies, with their corresponding errors. Early-type galaxies are shown in orange and late-type galaxies in blue. Orange and blue dash-dotted horizontal lines indicate the mean fraction of early- and late-type galaxies using the corresponding morphology catalogue, while dotted horizontal lines indicate the mean fraction of early- and late-type galaxies using the DS18 morphology, 25.2\% and 74.8\%, respectively, for comparison.} We use bins of 0.10\,$r/R_{void}$. The fraction of early-type galaxies has been shifted by +0.01\,$r/R_{void}$ so that the error bars do not overlap.
\label{fig:morpdistcomp}
\end{center}
\end{figure*}

When looking at galaxies surrounding voids, \citet{2017ApJ...846L...4R} found that the fraction of early-type galaxies increases from $\rm r/R_{void}$\,=\,0.8 to the outer region of the void (up to $\rm r/R_{void}$\,=\,2.5), which include galaxies that may be outside of the voids, i.e., galaxies belonging to filaments and walls (specially for the most spherical voids). According to the morphology-density relation, the number of early-type galaxies is larger in higher density environments. 
When looking at the region inside voids, we found that the fractions of early- and late-type galaxies are nearly constant with respect to void-centric distance. However, the $\rm f_{Late-type}$ is slightly higher than the median value within 0.1\,<\,$\rm r/R_{void}$\,<\,0.6, conversely, the $\rm f_{Early-type}$ is slightly lower than the median value in that void-centric distance range. This excess of late-type galaxies is also observable when comparing the results with other morphological classifications (see Fig.~\ref{fig:morpdistcomp}), and when considering the volume-limited sample of void galaxies. This result might indicate that there is more cold gas falling into galaxies in the inner region of the voids, within $\rm r/R_{void}$\,$\lesssim$\,0.5, favouring star-formation activity and therefore a larger fraction of late-type galaxies with respect to early-type galaxies. Our results therefore complement previous studies. 

It is worth noting, that \citet{2017ApJ...846L...4R} used the GZ1 morphological catalogue. According to our analysis, the completeness of this catalogue is low, where more than 60\% of galaxies do not have morphology classification (as shown in Table~\ref{tab:morph-cat}), and this might introduce additional uncertainties in the results. To check whether the  results might be affected by the use of different morphological classifications, in Fig.~\ref{fig:morpdistcomp} we repeated the same analysis as in Fig.~\ref{fig:morpdist} but using the GZ1 and GZ2 catalogues. We also used the CI parameter to compare with the results in \cite{2023MNRAS.521..916R}, which claim that the concentration index of galaxies increases from the centre of the voids to a distance $\rm r/R_{void}$\,$\lesssim$\,0.8 (with mean CI parameter $\sim$\,2.30, considering uncertainties), from where the CI is nearly constant at a mean value of $\sim$\,2.5, up to $\rm r/R_{void}$\,=\,4. It is important to note that the profile they presented in their results does not show mean CI values greater than 2.6 (i.e., early-type galaxies). The results of the reanalysis of our data sample to compare with \citet{2023MNRAS.521..916R} are also presented in Fig~\ref{fig:morpdistcomp}. We found similar results as using the DS18 morphology catalogue. Although the fractions of early- and late-type galaxies in voids are different depending on the morphology catalogue used, the trends with the void-centric distance remain nearly constant when using the CI parameter. 

It is important to note that the sample of voids used by \citet{2017ApJ...846L...4R} and \citet{2023MNRAS.521..916R} are also based on SDSS data, with void selected using volume-limited samples of galaxies, but using different void finder algorithms. In addition, \citet{2017ApJ...846L...4R} used the catalogue of voids compiled by \citet{2012ApJ...744...82V}, which provided a catalogue of galaxies in shells around voids, not inside voids. Since the catalogue of \citet{2012MNRAS.421..926P} only contains galaxies within voids, we cannot extend the sample to also cover galaxies around voids to perform a qualitative comparison, and therefore the results of this work are presented as a complement to previous studies.

% Figura 7__________________________________________________________________
% __________________________________________________________________

\subsection{Morphology of void galaxies as a function of the void size}\label{sec:dis-galmorphosize}

We did not find a significant difference in the fraction of early- and late-type galaxies in relation with the size of the void where they belong (see Fig.~\ref{fig:void-size}). As shown in Table~\ref{tab:void-size}, the fraction of early-type galaxies in small and in large voids is similar to the fraction measured considering the galaxies within all the voids in the Voids174 sample ($\rm f_{Early-type}$\,$\sim$\,25\%). Our results complement the results in \citet{2017ApJ...846L...4R}, who found that the surroundings of small voids host an excess of early-type galaxies and a lack of late-type galaxies, relative to their larger counterparts, which are preferably surrounded by late-type galaxies according to these authors. 
As we pointed out in Sect.~\ref{sec:dis-galmorphodistance}, \citet{2017ApJ...846L...4R} studied the dependence of the fraction of early-type galaxies as a function of void size, in the surrounding of voids, that is, in the outermost zone of the voids up to $\rm r/R_{void}$\,=\,2.5, which might also include galaxies belonging to filaments and walls. In our study, we focused on galaxies within voids, avoiding the walls surrounding them. Note that the small voids in \citet{2017ApJ...846L...4R} are much smaller (median value in the range 11.3-11.5\,$h^{-1}$\,Mpc) than the smallest void in the Voids174 sample. The smallest void in the Voids174 sample has an effective radius 11.6$h^{-1}$\,Mpc, therefore all the galaxies of our voids sample belong to large voids according to their classification. Note that, by definition, the VoidFinder algorithm in the Pan+12 void catalogue applied a cut-off of 10\,$h^{-1}$\,Mpc for the minimum radius of a void region, as explained in Sect.~\ref{sec:voidgal}.

As can be observed in Fig.~\ref{fig:void-size}, there is a large scatter around the global $\rm f_{Early-type}$ in the Voids174 sample within small (left of the dashed-red vertical line) and large (right of the dashed-red vertical line) voids. In particular, the scatter is larger (up to 0.4 in relative number of galaxies) in voids with a lower number of galaxies (indicated by smaller circles) than in voids with more galaxies (indicated by larger circles). This trend holds throughout the range of void sizes. The scatter in the plot is expected since, statistically, the estimation of the fraction of early- and late-galaxies might be affected when there is a low number of galaxies, showing extreme values of the fractions. In addition, as mentioned in Sect.~\ref{sec:res-galmorphosize} the scatter has a dependence with the redshift of the void where, at a given void size, the $\rm f_{Early-type}$ increases with redshift. We have checked that our results are consistent at different redshift bins and also considering an absolute magnitude volume-limited sample. The dispersion observed in Fig.~\ref{fig:void-size} together with the trend observed in the $\rm f_{Early-type}$ for voids with a low number of galaxies, motivates the analysis of the morphology of galaxies located in voids in terms of a new physical parameter for voids: the number-density of galaxies ($\rm \rho_N$), defined as the number of galaxies inhabiting a void ($\rm n_{galaxies}$), divided by the total volume of the void: 

\begin{equation}
    \rm \rho_N = \frac{n_{galaxies}}{4/3\,\pi\,R_{Void}^3} \quad,
\end{equation}
in units of $\rm gal/(h^{-1}\,Mpc)^3$, where $\rm n_{galaxies}$ is the number of galaxies in the void and $\rm R_{Void}$ is the effective radius of the void. Therefore, the galaxy number-density is a statistically robust parameter with low number of void galaxies, since it not only considers how many galaxies are in the void, but also the projected distance to their near neighbour galaxies, according to the size of the void, so it provides a better comprehension of the local environment of the galaxies within the void. Then, in voids with lower number-density, their galaxies are more isolated, even if the void is small if the projected separations between galaxies are large. In addition, the mean dispersion in the fraction of early- and late-type galaxies when using the number-density is smaller (0.05 in units of fraction of galaxies) than when considering the void size (0.07 in units of fraction of galaxies).

\begin{figure*}
\centering
% [trim={left bottom right top},clip]
\includegraphics[width=1\columnwidth, trim={0 0 1cm 1cm},clip]{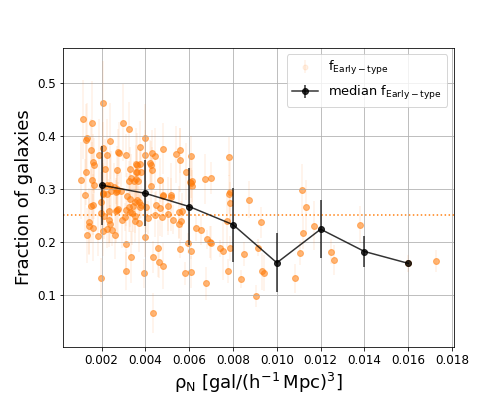}
\includegraphics[width=1\columnwidth, trim={0 0 1cm 1cm},clip]{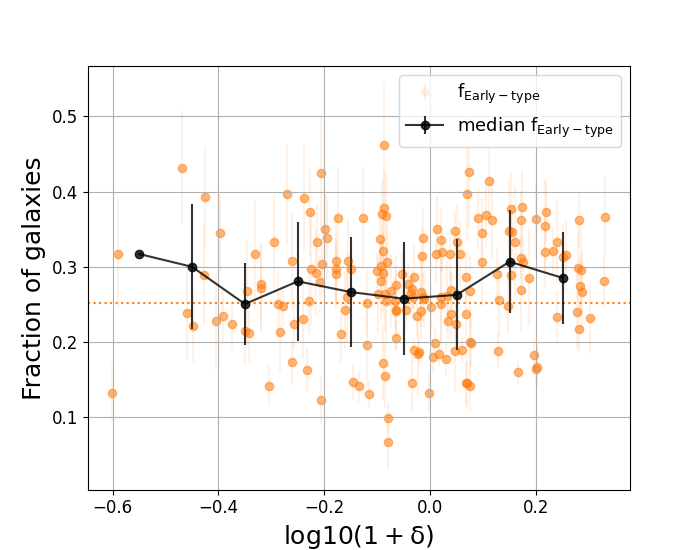}\\
\includegraphics[width=1\columnwidth, trim={0 0 1cm 1cm},clip]{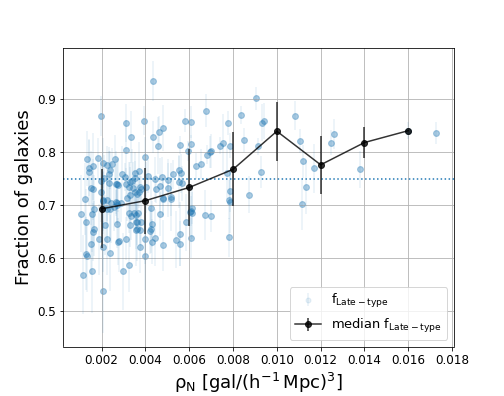}
\includegraphics[width=1\columnwidth, trim={0 0 1cm 1cm},clip]{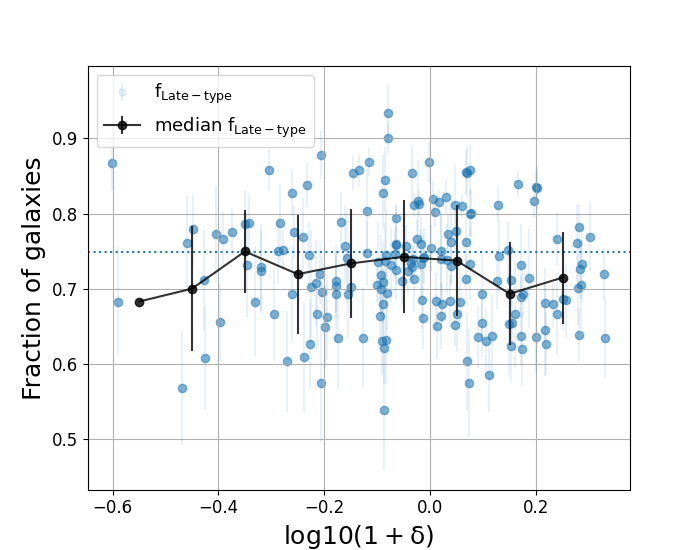}
\caption{Fraction of early- (upper panels) and late-type (lower panels) galaxies in each void as a function of its galaxy number-density (left panels) and the overdensity-corrected number-density (right panels). Each data point corresponds to the fraction of early-type galaxies (orange), and the fraction of late-type galaxies (blue), in each void. The black points and error bars indicate the median values, and corresponding 1$\sigma$, dispersion in eight bins of $\rm \rho_N$~=~0.002\,$\rm gal/(h^{-1}\,Mpc)^3$ (left) and nine bins of $\rm \rm log10(1 + \delta)~=~0.1$ (right). The orange and blue dotted horizontal line indicates the mean fraction of early- and late-type galaxies, 25.2\% and 74.8\% respectively, in the Voids174 sample, as presented in Table~\ref{tab:morph-cat}. 
}
\label{fig:void-den}
\end{figure*}

When exploring the dependency of the morphology of the void galaxies with the density of galaxies in the void, we initially find a decreasing trend for early-type galaxies, and an increasing trend with late-type galaxies. As shown in the left-hand panels in Fig.~\ref{fig:void-den}, the fraction of early-type galaxies in the less dense voids is about 5\% higher than the mean fraction in the Void174 sample, and decreases with the density of the void. Conversely, the fraction of late-type galaxies increases with increasing the number-density of galaxies. However, these trends are not present when considering the absolute magnitude volume-limited sample. This is because there is a dependency on the galaxy number-density with the redshift due to the Malmquist bias. Using a volume-limited sample of galaxies we find no dependency of the morphology of the void galaxies with the void galaxy number-density. However the galaxy number-density shows values concentrated between 0.0005 and 0.0040~$\rm [gal/(h^{-1}\,Mpc)^3]$ due to the low number of galaxies considered. This introduces a large scatter in the fraction of early-type galaxies. In addition, the volume-limited sample of void galaxies contains only galaxies brighter than $M_r$\,=\,$-$20.09, which favours the inclusion of early-type galaxies (which are typically brighter), under-representing faint galaxies for the voids at higher redshift. This is particularly critical in voids, considering that void galaxies are, in general, less massive and bluer than galaxies in denser environments. For these reasons, instead of analysing a volume-limited sample, we have chosen to apply overdensity corrections to the galaxy number-density for the full sample, which removes its dependency with the redshift without loosing galaxies. We adapt the methodology in \citet{2015MNRAS.451..660E} to estimate the overdensity for the void galaxy number-density, $\delta$, defined as:

\begin{equation}
    \rm \delta = \frac{\rho_N - \overline{\rho_N}}{\overline{\rho_N}}   \quad,
\end{equation}
where $\rm \overline{\rho_N}$ is the mean galaxy number-density for voids within redshift bins of z\,=\,0.01. The overdensity-corrected void galaxy number-density is then expressed in the form of $\rm \log_{10}(1 + \delta)$. As shown in the right-hand panels of Fig.~\ref{fig:void-den}, the fraction of early- and late-type galaxies has no dependency with the overdensity-corrected number-density of the void.

In summary, we found that the density of galaxies in the void do not has an impact on the morphology of the void galaxies. Thus, void galaxies do not follow the morphology-density relation within the void environment. The interpretation could be that the galaxy number-densities within voids are so low that the threshold to trigger the morphology-density relation is not reached yet. Specifically, the physical processes \citep[secular evolution, mergers, interactions, ram pressure stripping, etc,][]{2018A&A...620A.113A} responsible for the evolution from late towards earlier types (such as external environmental quenching) are not sufficiently effective in voids or so slow (internal secular quenching) that their contributions do not appear in the morphology-density relation.

% Figura 9__________________________________________________________________
% Figura 8__________________________________________________________________

%__________________________________________________________________
%% 6. Summary and Conclussions %%%%%%%%%%%%%%%%%%%%%%
\section{Summary and conclusions}\label{sec:sum}

In this work we investigate the effect of the Universe large-scale environment on the evolution of galaxies, by analysing the morphology of galaxies inhabiting cosmic voids. In particular, we study the morphology of galaxies as a function of the void-
centric distance and the void size. 

The sample of void galaxies that we use in this study is based on the catalogue of cosmic voids and void galaxies compiled by \citet{2012MNRAS.421..926P} using photometric and spectroscopic data from the SDSS-DR7. We select galaxies in the redshift range 0.005\,$\leq$\,$z$\,$\leq$\,0.080 to reduce uncertainties in the visual morphological classification due to poor spatial resolution. This leaves us with 26864 void galaxies in 174 voids. The size of the voids ($R_{void}$) is measured using the effective radius (radius of voids when they are spherical or, in the case where the void is not spherical, the radius of the sphere that has the same volume as the non-spherical void). We use the median value of the $R_{void}$ in \citet{2012MNRAS.421..926P} ($17.83\,h^{-1}$\,Mpc) to separate our sample into small (61 voids, containing 6285 void galaxies) and large (113 voids, containing 20579 void galaxies) voids.

We classify galaxies into early- and late-type  using the morphology classification by \cite{2018MNRAS.476.3661D}, who combined accurate existing visual classification catalogues with machine learning techniques, providing the largest and most accurate morphological catalogue up to date for galaxies in the SDSS. We found that 6335 void galaxies (25.2\%) are classified as early-type (E and S0), while 18839 void galaxies (74.8\%) are classified as late-type (Sa, Sb, Sc, Sd). We obtain consistent results when using other morphology classifications. We analysed the morphological distributions of these galaxies inside the voids by exploring how the galaxies are distributed within the voids as a function of the normalised void-centric distance, and how the galaxies are distributed as a function of the size and the number-density of galaxies in voids, defined as the number of galaxies inhabiting a void, divided by the total volume of the void.

Our main findings are the following:

\begin{itemize}
\item In agreement with previous studies, late-type is the predominant morphology for galaxies in voids. The morphology of galaxies in the Universe depends on the density of the environment: the low-density regions, known as voids, present a higher abundance of late-type galaxies than the high-density areas, such as clusters and filaments. This result is consistent with the morphology-density relation.

\item Although the number of late-type galaxies is larger than the number of early-type galaxies in voids, their distribution as a function of the normalised void-centric distance is similar. The fraction of early- and late-type galaxies are generally constant in the inner part of the voids up to a normalised distance of $\sim$1.2 ($\sim$75\% late- and $\sim$25\% early-types). We do not find either any dependence of the fraction of early- and late-type galaxies with respect to the size of the voids. Therefore, the morphology of void galaxies is not affected by their location within the void they reside, independently of the size of the void.

\item In addition, there is no dependence of the fraction of early- and late-type galaxies with respect to the number-density of galaxies in voids. Therefore, we find no difference between voids with lower or higher volume number-density of galaxies. 
\end{itemize}

Our results complement previous studies. \cite{2023MNRAS.521..916R} found that the concentration index of galaxies increases from the centre of the voids to a void-centric distance $\rm r/R_{void}$\,$\lesssim$\,0.8. In addition, \citet{2017ApJ...846L...4R} found that the fraction of early-type galaxies increases from $\rm r/R_{void}$\,=\,0.8 to the outer region of the void (up to $\rm r/R_{void}$\,=\,2.5), while we found that the fractions of early- and late-type galaxies are nearly constant with respect to the void-centric distance within the void, with a moderate excess of late-type galaxies within 0.1\,<\,$\rm r/R_{void}$\,<\,0.6. This might indicate that there is more cold gas falling into galaxies in the inner ($\rm r/R_{void}$\,$\lesssim$\,0.5) region of the voids. \citet{2017ApJ...846L...4R} also found that the surroundings of small voids host an excess of early-type galaxies and a lack of late-type galaxies, relative to their larger counterparts. In contrast, large voids are preferably surrounded by late-type galaxies according to these authors.

We conclude that, overall, galaxies in voids follow the morphology-density relation, in the sense that the majority of the galaxies in voids (the most under-dense large-scale environments) are late-type galaxies. However, we find no difference between voids with lower or higher volume number-density of galaxies: the fraction of early- and late-type galaxies do not depend on the density of the voids. The physical processes responsible for the evolution from late towards earlier types (such as external environmental quenching) are not sufficiently effective in voids or so slow (internal secular quenching) that their contributions do not appear in the morphology-density relation.

\begin{acknowledgements}
% UGR
The authors thank the anonymous referee for the thorough reading and constructive feedback. This paper is partially based on data obtained by the CAVITY project, funded by the Spanish Ministry of Science and Innovation under grant PID2020-113689GB-I00 as well as by Consejería de Universidad, Investigación e Innovación and Gobierno de España and Unión Europea - NextGenerationEU through grant AST22\_4.4.
We acknowledge financial support by the research project PID2020-114414GB-I00, financed by MCIN/AEI/10.13039/501100011033, the project A-FQM-510-UGR20 financed from FEDER/Junta de Andaluc\'ia-Consejer\'ia de Transforamción Económica, Industria, Conocimiento y Universidades/Proyecto and by the grants P20\_00334 and FQM108, financed by the Junta de Andalucía (Spain). 
% Mamen
M.A-F. acknowledges support from ANID FONDECYT iniciaci\'on project 11200107 and the Emergia program (EMERGIA20\_38888) from Consejer\'ia de Universidad, Investigaci\'on e Innovaci\'on de la Junta de Andaluc\'ia. 
% Salva
S.D.P. acknowledges financial support from Juan de la Cierva Formaci\'on fellowship (FJC2021-047523-I) financed by MCIN/AEI/10.13039/501100011033 and by the European Union "NextGenerationEU"/PRTR, Ministerio de Econom\'ia y Competitividad under grant PID2019-107408GB-C44 and PID2022-136598NB-C32, and also from the Fonds de Recherche du Qu\'ebec - Nature et Technologies. 
% Dani
DE acknowledges support from a Beatriz Galindo senior fellowship (BG20/00224) from the Spanish Ministry of Science and Innovation.\\

Funding for SDSS-III has been provided by the Alfred P. Sloan Foundation, the Participating Institutions, the National Science Foundation, and the U.S. Department of Energy Office of Science. The SDSS-III web site is http://www.sdss3.org/. SDSS-III is managed by the Astrophysical Research Consortium for the Participating Institutions of the SDSS-III Collaboration including the University of Arizona, the Brazilian Participation Group, Brookhaven National Laboratory, Carnegie Mellon University, University of Florida, the French Participation Group, the German Participation Group, Harvard University, the Instituto de Astrofisica de Canarias, the Michigan State/Notre Dame/JINA Participation Group, Johns Hopkins University, Lawrence Berkeley National Laboratory, Max Planck Institute for Astrophysics, Max Planck Institute for Extraterrestrial Physics, New Mexico State University, New York University, Ohio State University, Pennsylvania State University, University of Portsmouth, Princeton University, the Spanish Participation Group, University of Tokyo, University of Utah, Vanderbilt University, University of Virginia, University of Washington, and Yale University.\\

This research made use of Astropy, a community-developed core Python (http://www.python.org) package for Astronomy \citep{2013A&A...558A..33A, 2018AJ....156..123A}; ipython \citep{PER-GRA:2007}; matplotlib \citep{Hunter:2007}; SciPy, a collection of open source software for scientific computing in Python \citep{2020SciPy-NMeth}; pandas, an open source data analysis and manipulation tool \citep{reback2020pandas, mckinney-proc-scipy-2010}; and NumPy, a structure for efficient numerical computation \citep{2011CSE....13b..22V}.%; and Uncertainties: a Python package for calculations with uncertainties, Eric O. Lebigot, \url{http://pythonhosted.org/uncertainties/}.
\end{acknowledgements}

   \bibliographystyle{aa} % style aa.bst
   \bibliography{refs} % your references Yourfile.bib

\appendix

%% Figura A1__________________________________________________________________
%\begin{figure*}
%\centering
%% [trim={left bottom right top},clip]
%\includegraphics[width=1\textwidth, trim={1cm 0 1cm 1cm},clip]{Figures/frac_early_ngalvoids_174_240503.png} \\
%\includegraphics[width=1\textwidth, trim={1cm 0 1cm 1cm},clip]{Figures/frac_late_ngalvoids_174_240503.png}
%\caption{Fraction of early- (upper panel) and late-type (lower panel) galaxies in each void as a function of the number of galaxies in each void. Each data point corresponds to the fraction of early-type galaxies (orange), and the fraction of late-type galaxies (blue), in each void. The size of the circle around each data point is proportional to the size of each void (measured by their effective radius). The orange and blue dotted horizontal line indicates the mean fraction of early- and late-type galaxies, 25.2\% and 74.8\% respectively, in the Voids174 sample, as presented in Table~\ref{tab:morph-cat}.}
%\label{fig:void-ngal}
%\end{figure*}
%% Figura A1__________________________________________________________________

\end{document}